# Engineering Plasmons in Oxide/Graphene Heterostructures via Interfacial Charge Transfer

Yuanchen Chi[1,2], Dongxu Di[1], Michael Fralaide[1], Jigang Wang[1,2], Zhe Fei[1,2]

[1]Department of Physics and Astronomy, Iowa State University, Ames, Iowa 50011, USA
[2]Ames National Laboratory, U.S. Department of Energy, Iowa State University, Ames, Iowa 50011, USA

*Correspondence: (Z.F.) zfei@iastate.edu

**Keywords**

Graphene plasmons, Interfacial charge transfer, Oxide/graphene heterostructures, Near-field nano-optics, s-SNOM

**Abstract**

Interfacial charge transfer provides an effective route for tailoring the optical and electronic properties of two-dimensional materials. Here, we investigate infrared surface plasmon polaritons in oxide/graphene heterostructures using scattering-type scanning near-field optical microscopy. Ultrathin oxide overlayers deposited by physical vapor deposition enable systematic engineering of graphene plasmons through interfacial charge redistribution. $MoO_x$ strongly enhances the plasmonic response, producing a longer plasmon wavelength, stronger fringe contrast, and reduced damping, whereas a subsequently deposited $ZnO_x$ overlayer partially reverses these changes. Energy-dependent nano-infrared imaging combined with quantitative modeling reveals an increased graphene carrier density and the resulting modification of the plasmon dispersion. Thickness-dependent measurements show a rapid increase in charge-transfer doping at sub-nanometer $MoO_x$ thicknesses, followed by a weaker long-range contribution at larger overlayer thicknesses. Electrostatic gating further modulates the carrier density and produces a nonlinear response consistent with gate-dependent interfacial charge redistribution. In addition, an approximately 3-nm-thick $MoO_x$ overlayer stabilizes the plasmonic response for at least seven months under ambient conditions. These results establish oxide/graphene heterostructures as a robust platform compatible with scalable fabrication, providing a pathway toward stable and tunable infrared nanophotonic and optoelectronic devices.

## 1. Introduction

Graphene plasmonics has attracted considerable interest because of graphene's ability to support highly confined and electrically tunable surface plasmon polaritons (SPPs) across a broad spectral range from the terahertz (THz) to the infrared (IR) [1–4]. These properties enable strong light–matter interactions at deep-subwavelength length scales and make graphene an attractive platform for nanophotonic and optoelectronic applications [2,5–8]. A central challenge in graphene plasmonics, however, is achieving and maintaining carrier densities that are sufficiently high and stable to support strong plasmonic responses over targeted spectral ranges. Although electrostatic gating provides reversible control of the carrier density, its practical implementation requires gate electrodes and dielectric layers and can be constrained by device geometry, dielectric breakdown, and charge trapping or hysteresis. Therefore, scalable approaches that provide high and stable carrier densities without continuous electrical bias are desirable for practical graphene plasmonic devices.

Charge-transfer doping provides an attractive route for controlling the graphene carrier density. When graphene is brought into contact with a material of different work function, interfacial charge redistribution can occur until electrochemical equilibrium is established, resulting in electron or hole doping without requiring covalent modification of the graphene lattice [9-11]. Various charge-transfer strategies have been explored, including the intercalation of atomic and molecular species [12–19] and molecular adsorption [20–24]. Although these approaches can effectively tune the carrier density, intercalation often requires stringent processing conditions and may alter the intrinsic electronic properties of graphene, whereas molecular adsorption can be susceptible to desorption and ambient-induced variability. Metal oxides provide a promising alternative because their work functions span a broad range and ultrathin oxide films can be deposited directly onto graphene. Depending on the oxide composition and interfacial energy alignment, such layers can increase or decrease the carrier density of graphene. These considerations motivate the development of solid-state charge-transfer platforms that combine strong doping, environmental stability, and fabrication scalability.

In this work, we investigate oxide/graphene heterostructures as a platform for engineering graphene plasmons through interfacial charge transfer. Using physical vapor deposition (PVD), we form ultrathin $MoO_x$ and $ZnO_x$ overlayers with controlled nominal thicknesses. Compared with previously reported oxide-formation approaches based on the post-growth oxidation of layered materials [25], PVD provides a direct and scalable route for fabricating oxide/graphene heterostructures. This approach enables systematic variation of the oxide composition and thickness while preserving compatibility with electrostatic gating. Using nano-infrared imaging and quantitative modeling, we determine how these parameters influence the graphene carrier density, plasmon wavelength, and damping. Our results establish oxide deposition as an effective strategy for stable and tunable graphene plasmonics, with the $MoO_x$ overlayer additionally providing long-term environmental stabilization.

## 2. Results and Discussion

To investigate plasmonic responses in oxide/graphene heterostructures, we employ scattering-type scanning near-field optical microscopy (s-SNOM) integrated with a tapping-mode atomic force microscope (AFM), enabling simultaneous acquisition of near-field optical signals and surface topography. Figure 1a illustrates the experimental configuration. The metallic tip is illuminated by a continuous-wave $CO_2$ laser operating at discrete IR wavelengths $\lambda_0$ = 9.2–10.8 μm (115–135 meV). Graphene samples are prepared by mechanical exfoliation onto $SiO_2$/Si substrates, followed by deposition of $MoO_x$ and $ZnO_x$ overlayers via physical vapor deposition (PVD). The oxide thickness is monitored during deposition and subsequently confirmed by AFM measurements. All s-SNOM measurements were performed under ambient conditions.

In Fig. 1b-d, we present the nano-IR images of a bare graphene sample (G), a $MoO_x$/graphene heterostructure (M/G), and a $ZnO_x$/$MoO_x$/graphene heterostructure (Z/M/G). The graphene sample is produced by mechanical exfoliation and then transferred to the standard $SiO_2$/Si wafers. The heterostructures were produced by subsequent deposition of $MoO_x$ and $ZnO_x$ on the same graphene sample. The experimental observable plotted in the images is the third-harmonic near-field amplitude [26]. The thickness of both the $MoO_x$ and $ZnO_x$ overlayer is about 0.5 nm. The IR excitation energy is set to be $E$ = 115 meV, which is away from the strong optical phonon of $SiO_2$ (centered at around 140 meV), so it is ideal for imaging graphene SPPs. In all three nano-IR images, one can see bright fringes parallel to the edge of the sample (marked with white dashed curves). As introduced in previous works, these fringes are generated due to the interference of tip-launched SPPs and those reflected by the sample edges [3,4]. From the nano-IR images, one can see that the SPP fringe in bare graphene is the weakest. Adding a thin $MoO_x$ layer to graphene strongly enhances the fringes: both the fringe intensity and width are increased. Further addition of a thin $ZnO_x$ layer significantly reduces the fringe intensity and width.

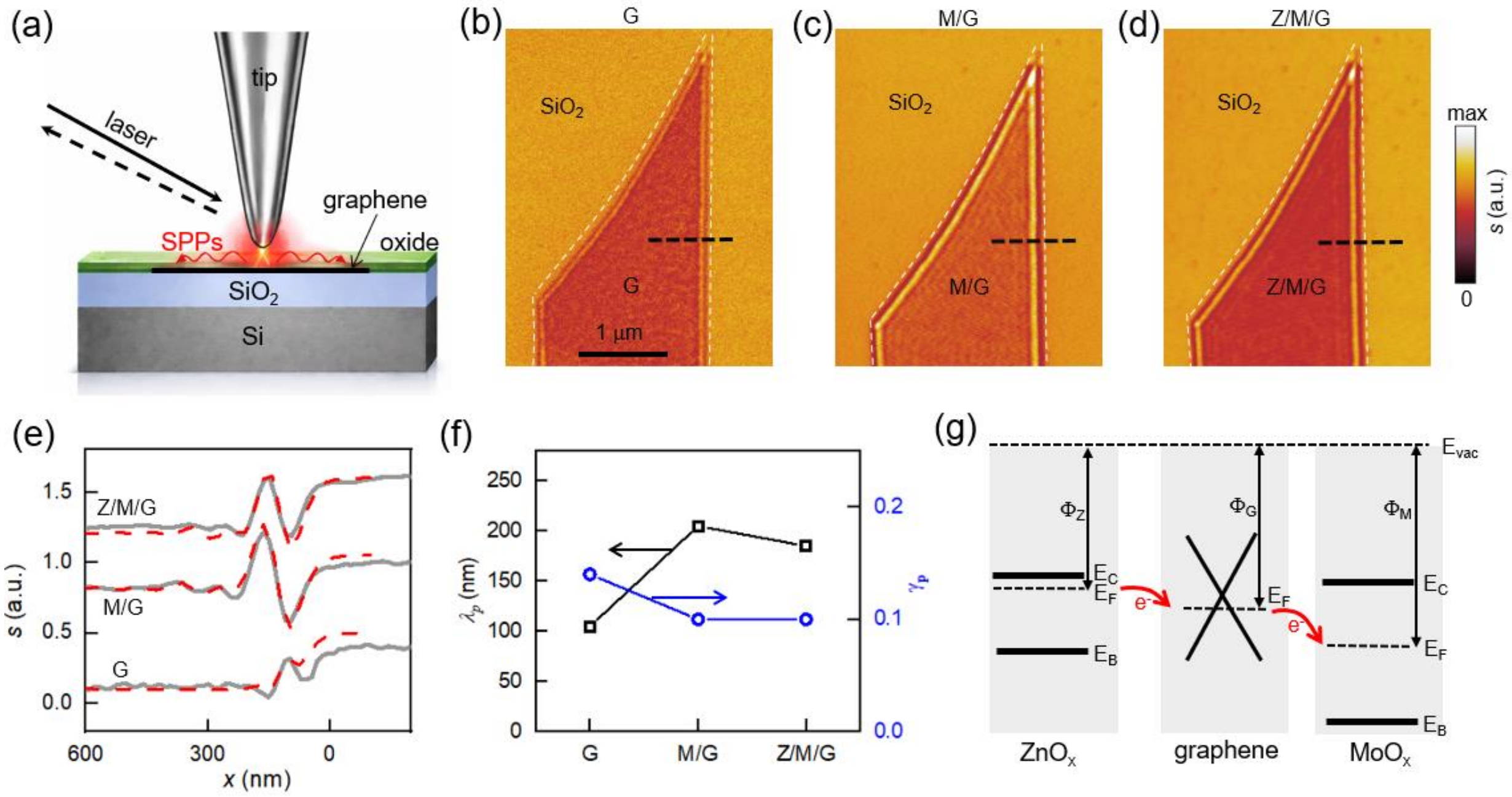

**Figure 1.** (a) Schematic of the nano-IR imaging measurement of an oxide/graphene heterostructure. (b-d) Nano-IR images of bare graphene (G), $MoO_x$/graphene (M/G) heterostructure, and $ZnO_x$/$MoO_x$/graphene (Z/M/G) heterostructure, respectively. The excitation laser energy is set to be 115 meV. (e) Line profiles perpendicular to the edges of the sample taken along the dashed lines in panels (b-d). (f) Extracted plasmon wavelength $\lambda_p$ (black squares) and damping rate $\gamma_p$ (blue circles) by fitting the profiles in panel (e). (g) Band alignment diagram illustrating charge transfer between graphene and the two types of oxides.

For quantitative analysis of the SPP fringes, we extract horizontal line profiles perpendicular to the sample edges from the nano-IR images (along the black dashed lines in Fig. 1b-d). We then fit the profiles with a quantitative s-SNOM model. In the model, we approximate the s-SNOM tip as a conducting spheroid. The total radiating dipole moments of the tip are calculated at various tip locations and tip-sample distances, based on which we compute the third-harmonic near-field amplitude and the fringe profiles. A detailed introduction about the s-SNOM model is given in a previous study [3]. Similar fringe analyses have been performed on all SPP imaging data in the work.

In the model, graphene or the heterostructure sample is modeled as an ultrathin plasmonic film with a complex plasmon wavevector ($q_p = q_1 + iq_2$). This parameter can fully describe all plasmonic responses of graphene and is directly related to its electronic properties. Under 2D and Drude approximation, $q_p$ can be written as [27,28]:

$$q_p \equiv q_1 + iq_2 \approx \frac{2\pi\varepsilon_0\kappa}{e^2 E_F} E(E + iE_\gamma). \quad [1]$$

Here, $E_F$ is the Fermi energy of graphene, $E_\gamma = \hbar\gamma$ is the charge scattering energy, and $\kappa = \kappa_1 + i\kappa_2$ is the effective dielectric function of the graphene environment. From $q_p$, we can directly obtain the plasmon wavelength ($\lambda_p = 2\pi/q_1$) and damping rate ($\gamma_p = q_2/q_1$). Based on Eq. 1, $\lambda_p$ can be written as:

$$\lambda_p = \frac{2\pi}{q_1} \approx \frac{e^2 E_F}{\varepsilon_0 \kappa E^2}. \quad [2]$$

Therefore, $\lambda_p$ is roughly proportional to $E_F$, so it is sensitively dependent on the charge-transfer doping. Note that the approximate formulas in Eqs. 1 and 2 are mainly for qualitative discussions. The plasmonic modeling and dispersion calculations shown in the work are based on rigorous calculations considering the entire $MoO_x$/graphene/$SiO_2$/Si multilayer structure (see Fig. 1a).

The experimental and modeling profiles are plotted in Fig. 1e, which show good agreement. Through the fitting, we were able to extract $q_p$, $\lambda_p$ and $\gamma_p$. The latter two were plotted in Fig. 1f. From Fig. 1f, one can see that $\lambda_p$ reaches 205 nm at $E$ = 115 meV after the deposition of $MoO_x$, which is approximately twice the plasmon wavelength of bare graphene (105 nm). After an additional deposition of $ZnO_x$, $\lambda_p$ drops from 205 nm to 185 nm. The result indicates that $MoO_x$ increases the doping of graphene while $ZnO_x$ slightly decreases the doping. The change of doping is due to the charge transfer between graphene and oxides. As illustrated in Fig. 1g, pristine graphene on $SiO_2$/Si substrates is initially hole doped. Its work function ($\Phi_G$ ~ 4.8 eV) is substantially lower than that of $MoO_x$ ($\Phi_M$ ~ 5.5-6.8 eV) [29,30], so graphene will transfer electrons to $MoO_x$, thus increasing its own hole doping. In contrast, the work function of $ZnO_x$ ($\Phi_Z$ ~ 4.2-4.6 eV) is slightly lower than hole-doped graphene [31-33], so it will transfer electrons to graphene and decrease its hole doping. In addition to $\lambda_p$, we also found that $\gamma_p$ drops after deposition of $MoO_x$ (see Fig. 1f). This is again due to the higher carrier density of graphene, which can screen the charged impurities around graphene, thus reducing the scattering rate. Similar ambipolar charge-transfer doping has also been seen when graphene is interfacing with $WO_x$ and $ZrO_x$ [25].

To construct the plasmon dispersion relationship $E(q_p)$, we performed nano-IR imaging of the M/G heterostructure sample at various excitation IR energies. As shown in Fig. 2a-f, the intensity and width of the SPP interference fringes at the sample edge decrease systematically with $E$. At $E$ = 129 meV, the plasmon fringes are barely seen. This is due to the strong coupling between graphene SPPs and optical phonons of $SiO_2$ at 140 meV, resulting in a significant plasmon mode flattening and damping [34]. After extracting and fitting the plasmon fringe profiles, we obtained $\lambda_p$ and $q_p$ at every IR energy, thus establishing the plasmon dispersion. The experimental dispersion is plotted in Fig. 2g as data points, which are overlaid on a theoretical dispersion colormap. The colormap is generated by computing the imaginary part of the $p$-polarized reflection coefficient $r_p(q, \omega)$. As introduced in previous literature [34], SPPs or other types of polaritons can be visualized in such colormaps as bright modes (marked with a blue curve in Fig. 2g) due to the divergence of $r_p$ as the mode emerges. By matching the experimental data points with dispersion colormap, we found the Fermi energy ($E_F$) of graphene in the M/G heterostructure is about 0.38 eV, which is approximately twice that of $E_F$ of bare graphene (~0.2 eV). Accordingly, the carrier density increases from $2.9\times10^{12}$ cm$^{-2}$ in the bare graphene to $1.1\times10^{13}$ cm$^{-2}$ in the M/G heterostructure. Due to the elevated doping, the plasmon dispersion shifts to lower momentum (from white curve to blue curve in Fig. 2g) resulting in a larger $\lambda_p$, which is consistent with Eq. 2.

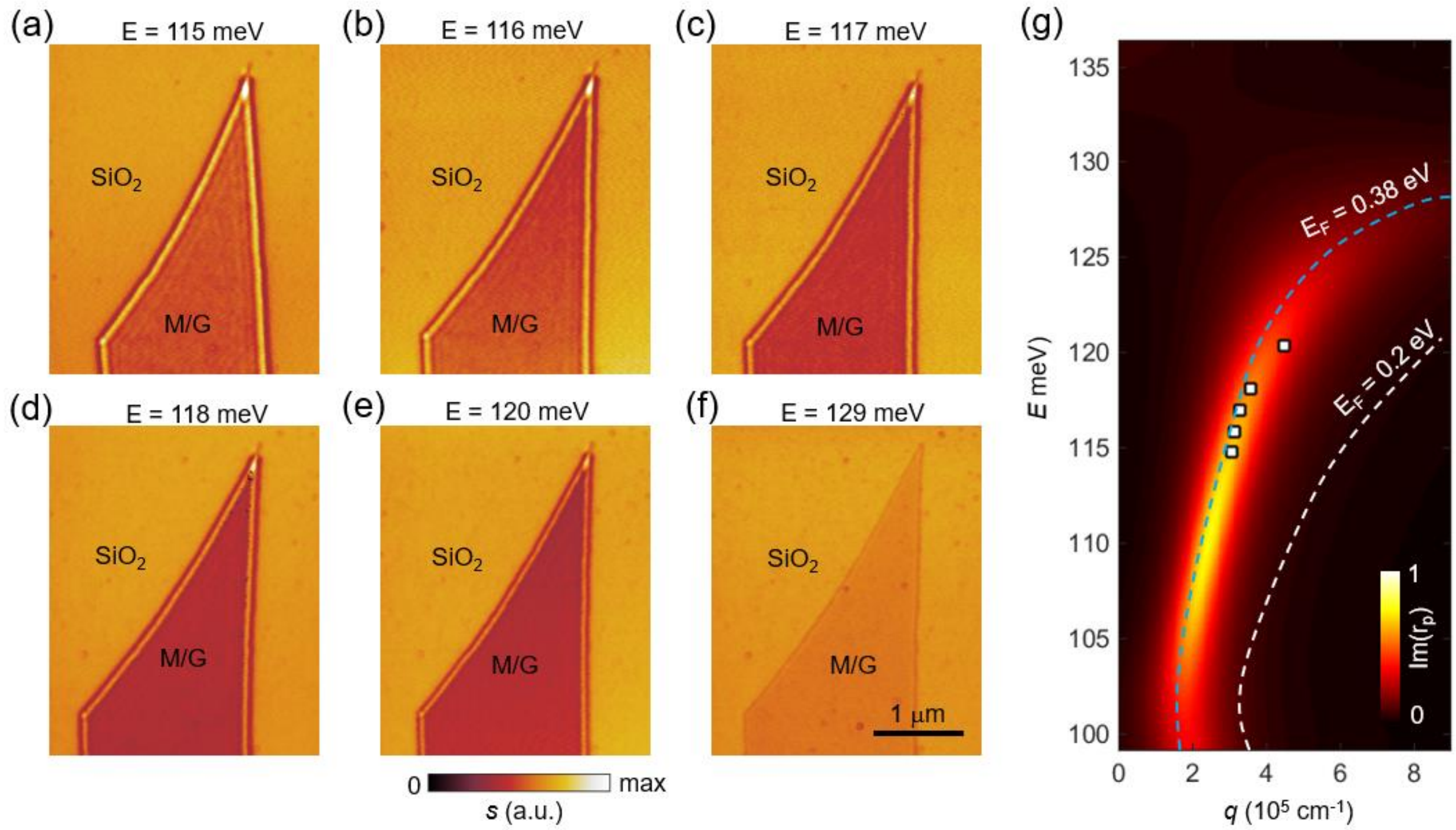

**Figure 2.** (a–f) Nano-IR images acquired at different excitation energies ($E$). (g) Plasmon dispersion relationship from both experimental data (white squares) and theoretical calculations (colormap).

The results in Figs. 1 and 2 demonstrate the effectiveness of charge-transfer doping by an ultrathin (~0.5 nm) $MoO_x$ overlayer in enhancing graphene SPPs. To further elucidate this mechanism, we perform systematic nano-IR imaging while using the $MoO_x$ thickness ($d$) as a tuning parameter. Specifically, $MoO_x$ is deposited on graphene in a stepwise manner, and after each deposition step, a nano-IR image is acquired to track the evolution of the SPP interference fringes. The images are plotted in Fig. 3a-d, where the plasmon fringes clearly evolve with $d$. The bare graphene sample studied here has an extremely weak plasmon fringe indicating low doping (Fig. 3a). After depositing 0.1 nm of $MoO_x$, we can see a clear plasmon fringe at the sample edge (Fig. 3b). In addition, there are clear signal inhomogeneities in the interior of the sample indicating that $MoO_x$ is not uniformly deposited. This is expected due to the limited amount of $MoO_x$ deposited. Further depositions ($d$ = 0.9 nm and 2 nm) make the interior more uniform, and the plasmon fringes become stronger and well separated from each other indicating higher doping (Fig. 3c,d). Thicker $MoO_x$ layers ($d$ = 6.5 nm and 10.1 nm) retain the broad plasmon fringe, but the fringe intensity drops significantly primarily due to the large tip-graphene distance (Fig. 3e,f).

By fitting the fringe profiles using quantitative modeling, we extract the plasmon wavelength $\lambda_p$ from the images in Fig. 3a–f and plot the results in Fig. 3g. The plasmon wavelength initially increases with the $MoO_x$ thickness $d$and reaches a maximum near $d \approx$ 2nm. Further $MoO_x$ deposition produces a slight decrease in $\lambda_p$, primarily because the thicker overlayer enhances dielectric screening. More specifically, the relatively thick $MoO_x$ layers of 6–10 nm increase the effective dielectric constant $\kappa$ of the graphene environment, resulting in a smaller $\lambda_p$, as indicated by Eq. 2. To separate the dielectric and electronic contributions, we use multilayer dispersion calculations that explicitly account for the $MoO_x$ thickness at each deposition step to determine the thickness-dependent Fermi energy $E_F(d)$and carrier density $n(d)$ (See Supporting Information). The resulting $n(d)$ curve, shown in Fig. 3h, exhibits two distinct regimes: a rapid increase for $d \leq$ 1nm, followed by much slower growth for $d >$ 1nm.

We describe this behavior using a phenomenological two-length-scale model,

$$n(d) = n_0 + A_1\left(1 - e^{-d/L_1}\right) + A_2\left(1 - e^{-d/L_2}\right),$$

where $n_0$ is the carrier density of bare graphene, and the two terms represent short- and longer-range contributions to the thickness-dependent charge redistribution. Fitting the thickness dependence yields effective characteristic lengths of $L_1 = 0.4 \pm 0.1$nm and $L_2 = 15 \pm 5$nm. The short length scale is consistent with atomic-scale interfacial coupling and the progressive coverage of graphene by $MoO_x$. The longer length scale is comparable to the estimated graphene-plasmon intensity decay length in $MoO_x$, $L_I \approx \lambda_p/4\pi \approx 16$ nm (see Supporting Information). This numerical similarity suggests that plasmon-assisted charge transfer may be one possible origin of the longer-range contribution [35,36]. Because the graphene-plasmon field extends vertically into the $MoO_x$ overlayer, the strength of such a process is expected to decay exponentially with distance from graphene. Other mechanisms, including defect- or trap-mediated charge redistribution, localized hopping transport, and electrostatic redistribution within $MoO_x$, may also contribute. However, the $MoO_x$ films exhibit no measurable electrical conductivity within our experimental sensitivity, suggesting that conventional charge transport through extended electronic states is unlikely to dominate the observed longer-range contribution.

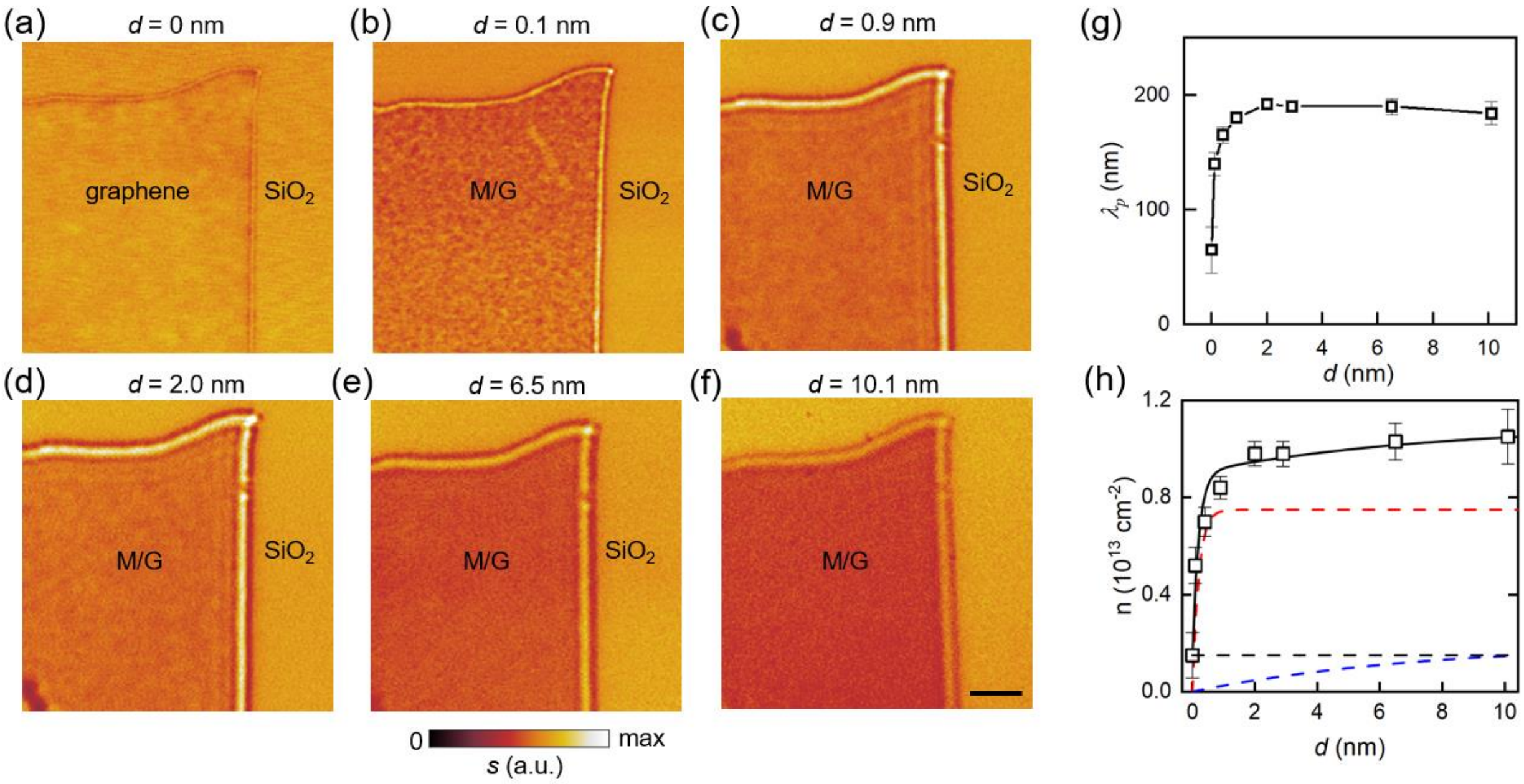

**Figure 3.** (a-f) Nano-IR images of bare graphene and $MoO_x$/graphene heterostructures with different thicknesses of $MoO_x$. (g,h) The extracted plasmon wavelength $\lambda_p$ and carrier density $n$ (data points) versus deposition thickness of $MoO_x$.

One advantage of graphene heterostructures fabricated with top-down methods (e.g., PVD or mechanical transfer) is the preserved capability of electrical gating. In comparison, intercalation [12-19] – the other method capable of producing stable charge-transfer graphene heterostructures – works mainly on few-layer graphene or graphite, so it might lose the capability or efficiency of electric gating due to the relatively large sample thickness. Nevertheless, the gating response of graphene in the presence of charge transfer is not fully understood. In Fig. 4a-f, we show the nano-IR imaging data of a $MoO_x$/graphene heterostructure device with back gating. Here the oxide thickness is about 0.5 nm. Clearly, the plasmon-fringe patterns evolve systematically with the applied gate voltage ($V_g$). By fitting the fringe profiles, we extracted gate-dependent $\lambda_p$ and carrier density ($n_G$) of graphene, which are plotted in Fig. 4g and Fig. 4h, respectively. Here, one can see that the carrier density reaches as high as $1.14 \times 10^{13}$ cm$^{-2}$ at $V_g$ = -100 V, corresponding to a Fermi energy of about 0.4 eV. Due to the high doping, $\lambda_p$ reaches 220 nm at $E$ = 115 meV. As we tuned $V_g$ towards positive voltages, we found that carrier density does not decrease linearly.

Instead, the decrease of carrier density slows down so $n_G(V_g)$ dependence curve bends upward. This is different from pristine graphene, where the electron density of graphene has approximately a linear dependence with $V_g$ (red dashed curve in Fig. 4h):

$$n_G(V_g) \approx \alpha_{ox}(V_g - V_0). \quad [3]$$

Here, $V_0$ is the charge neutrality voltage, and $\alpha_{ox} \approx 7.2 \times 10^{10}$ $cm^{-2}V^{-1}$ is the gate-coupling coefficient. For pristine graphene on $SiO_2$, $V_0$ is normally in the range from 10 - 40 V in our samples, so graphene is initially hole doped ($n_G < 0$) without gating.

After adding a thin $MoO_x$ layer to graphene, electrons transfer from graphene to $MoO_x$, shifting the charge-neutrality voltage, denoted $V_0'$, to a much higher positive value—possibly close to 100 V for the sample shown in Fig. 4. In addition, $MoO_x$ acquires an interfacial electron density of $n_M$ from graphene due to the charge transfer. Therefore, the gating equation becomes:

$$n_G(V_g) + n_M(V_g) \approx \alpha_{ox}(V_g - V_0'). \quad [4]$$

Note that $n_M$ is roughly proportional to the difference of work function of $MoO_x$ and graphene: $n_M \propto \Delta\Phi = \Phi_M - \Phi_G$). The work function of graphene (see Fig. 1g) can be written as:

$$\Phi_G = \Phi_G^0 \pm \hbar v_F \sqrt{\pi |n_G|}\,, \quad [5]$$

with "+" for hole doping and "–" for electron doping, and $\Phi_G^0 \approx 4.5\ eV$ for charge-neutral graphene. From Eqs. 4 and 5, it is clear $\Phi_G$ does not scale linearly with $V_g$, which explains why $n_G$ has a nonlinear relationship with $V_g$.

Back to the gating data shown in Fig. 4h, graphene is heavily hole-doped at $V_g$ = -100 V, so $\Phi_G \approx 4.9\ eV$ is larger than $\Phi_G^0$. As a result, $\Delta\Phi$ and the transferred electron density (namely $n_M$) will be smaller compared to those of charge-neutral graphene. As we tune $V_g$ towards positive voltages, graphene becomes less hole doped, so the charge transfer will become stronger. It will be even stronger if graphene switches to the electron doping regime at very large positive $V_g$. Such a gate-dependent charge transfer explains the unusual behaviors of $\lambda_p(V_g)$ and $n_G(V_g)$ shown in Fig. 4g,h, which are different compared to those of pristine graphene (red dashed curves).

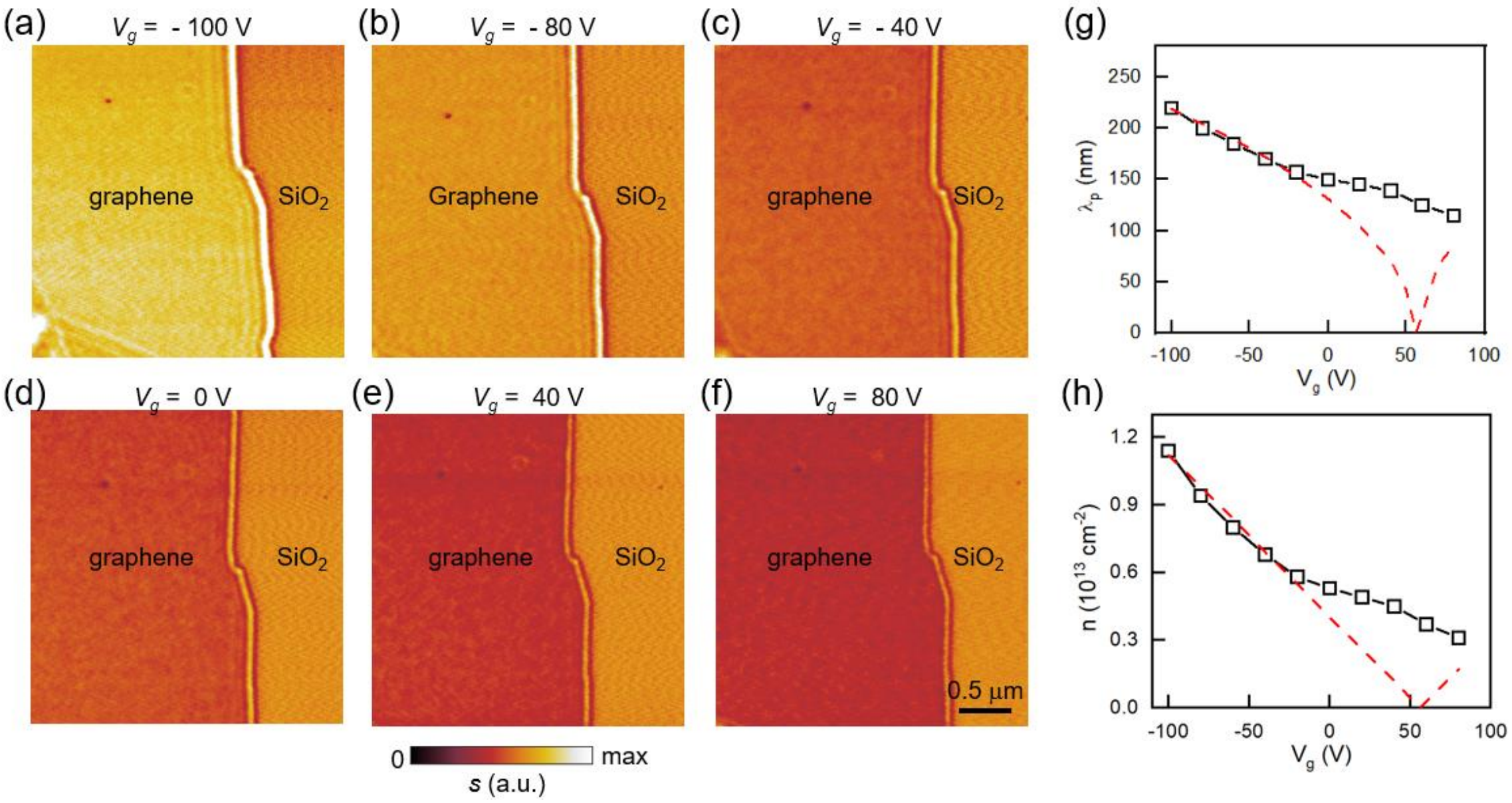

**Fig. 4.** (a-f) Nano-IR images of a $MoO_x$/graphene heterostructure acquired under various back-gate voltages $V_g$. (g,h) The extracted plasmon wavelength $\lambda_p$ and carrier density $n$ versus gate voltage ($V_g$). The red dashed curves in panels g and h are calculated dependence curves for pristine graphene based on Eqs. 2 and 3.

Finally, we emphasize an additional technical advantage of the oxide overlayer beyond charge-transfer doping and plasmon enhancement. The deposited $MoO_x$ acts as an effective passivation layer, protecting graphene from ambient degradation due to moisture and oxygen. Without encapsulation, the carrier density and plasmonic response of graphene can vary under ambient conditions because of adsorption, desorption, and other environmental effects. To evaluate the stability imparted by the oxide, we performed nano-IR imaging on a $MoO_x$/graphene heterostructure ($MoO_x$ thickness ≈ 3 nm) immediately after fabrication (Fig. 5a) and repeated the measurement after 7 months (Fig. 5b). The plasmon interference fringe patterns remain nearly unchanged over this period. For a quantitative comparison, we extract edge-normal line profiles and fit the fringe oscillations, as shown in Fig. 5c. The extracted plasmon wavelength $\lambda_p$ is about 193 nm initially and 197 nm after 7 months, indicating negligible degradation and confirming the long-term environmental stability provided by the $MoO_x$ overlayer.

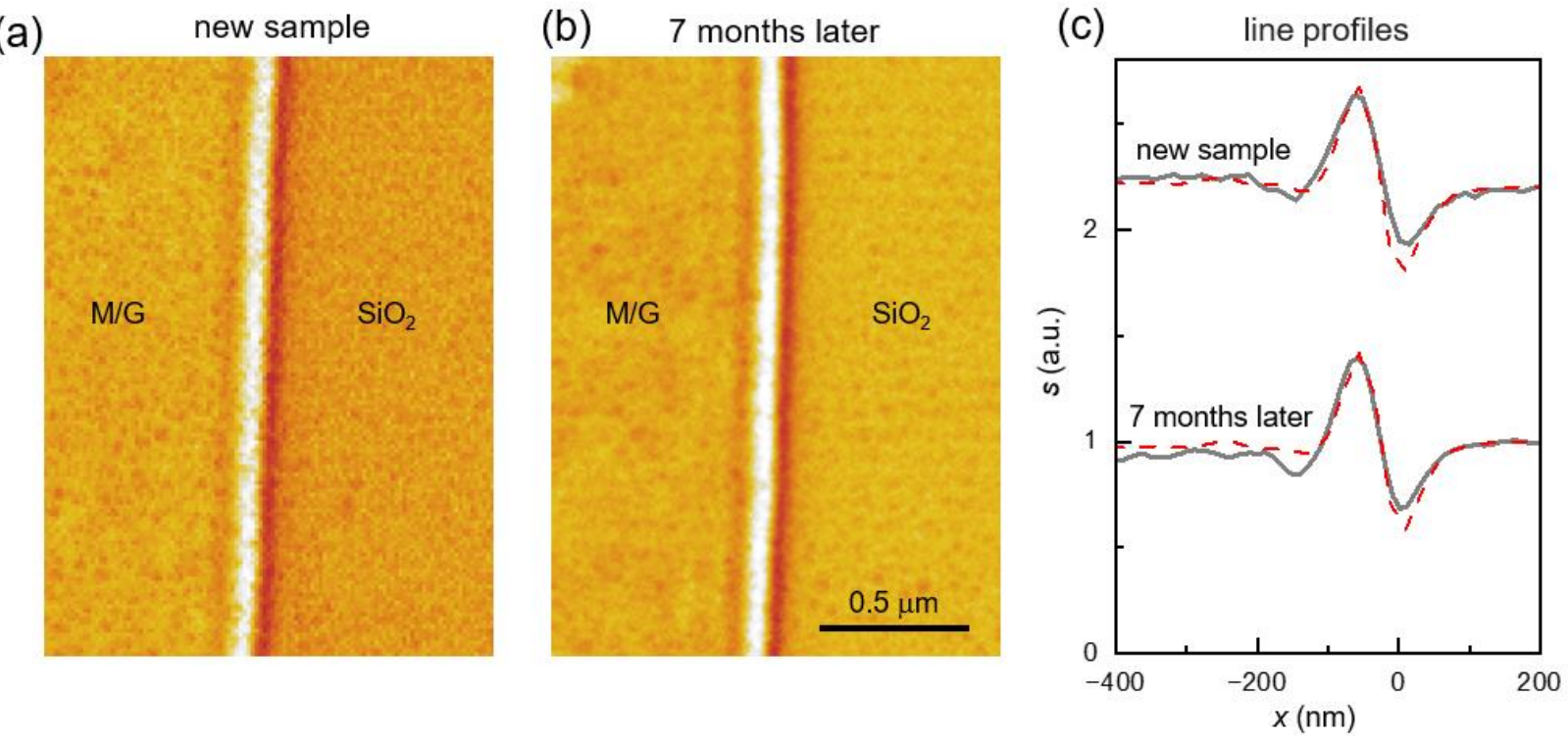

**Fig. 5.** (a,b) Nano-IR images of a $MoO_x$/graphene heterostructure acquired immediately after fabrication (a) and after 7 months of ambient exposure (b). (c) Line profiles (gray) extracted perpendicular to the SPP fringes in (a) and (b), together with the modeling profiles (red).

## 3. Conclusion

In summary, we systematically investigate SPPs in oxide/graphene heterostructures using nano-IR imaging, establishing oxide deposition as a versatile route to engineer graphene plasmons. By comparing $MoO_x$/graphene and $ZnO_x$/$MoO_x$/graphene, we demonstrate opposite-direction charge transfer, allowing the $MoO_x$-induced doping to be partially reversed by $ZnO_x$. Energy-dependent measurements resolve the plasmon dispersion, while thickness-dependent studies reveal a rapid short-range contribution followed by a weaker longer-range contribution. Electrostatic gating provides an active tuning knob that not only modulates carrier density but also dynamically tunes interfacial charge transfer via work function alignment. In addition, the oxide overlayer serves as an effective passivation layer, preserving plasmonic performance over months under ambient conditions. Beyond the specific material system, this work establishes oxide/graphene heterostructures as a scalable and robust platform for plasmonic engineering. The compatibility of PVD with large-area processing, for example with shadow-mask techniques [37], enables spatially controlled oxide patterning and programmable plasmonic responses. This capability opens a new pathway toward reconfigurable plasmonic metasurfaces. Overall, our results provide a practical platform for realizing stable, tunable nanophotonic and optoelectronic devices based on graphene plasmons.

**Supporting Information**

Supporting Information is available from the Wiley Online Library or from the author.

**Acknowledgements**

This work was supported by the National Science Foundation under Grant No. DMR1945560. The nano-infrared imaging experiments were partially supported by Ames National Laboratory. Ames National Laboratory is operated for the U.S. Department of Energy by Iowa State University under Grant No. DE-AC02-07CH11358.

**Conflict of Interest**

The authors declare no conflict of interest.

**Supporting Information**

**Engineering Plasmons in Oxide/Graphene Heterostructures via Interfacial Charge Transfer**

Yuanchen Chi[1,2], Dongxu Di[1], Michael Fralaide[1], Jigang Wang[1,2], Zhe Fei[1,2]

[1]Department of Physics and Astronomy, Iowa State University, Ames, Iowa 50011, USA
[2]Ames National Laboratory, U.S. Department of Energy, Iowa State University, Ames, Iowa 50011, USA

*Author to whom correspondence should be addressed: (Z.F.) zfei@iastate.edu

## S1. Sample and device fabrication

Graphene flakes were mechanically exfoliated from natural graphite onto $SiO_2$/Si substrates with a 300-nm-thick $SiO_2$ layer. Oxide overlayers were subsequently deposited using a physical vapor deposition (PVD) system under ultrahigh-vacuum conditions. $MoO_x$ films were deposited using $MoO_3$ source material, whereas $ZnO_x$ films were deposited using ZnO source material. During deposition, a resistive heating current of typically 20–30 A was applied to evaporate the source materials. The deposition rate was maintained at approximately 0.01 nm $s^{-1}$, and the film thickness was controlled by adjusting the deposition time. The deposited thickness was monitored in situ using a capacitance-based thickness monitor. Prior to sample fabrication, the thickness measurements were calibrated against atomic force microscopy (AFM) measurements of deposited films. For gating experiments, electrical contacts to graphene were fabricated by directly applying silver or graphite paste. No lithographic processing was employed at any stage of sample fabrication to minimize contamination and preserve the intrinsic properties of graphene and oxide/graphene heterostructures.

## S2. Properties of the MoOx films

Thermally evaporated $MoO_x$ films prepared under vacuum have reported stoichiometric ratios in the range x=2.58–2.96 [1–3]. Before the s-SNOM measurements, the oxide/graphene samples were exposed to ambient air for at least 2–3 h, which is expected to further oxidize the films. Electrical measurements detected no measurable conductivity within our experimental sensitivity, consistent with a highly oxidized, nearly stoichiometric $MoO_x$ composition with $x$ close to 3 [4]. If $x$ drops below 3.0 instead, the film will become more conductive. This is because removing oxygen normally releases electrons and reduces Mo: Mo 6+ →Mo 5+ →Mo 4+. The resulting Mo 4d-derived defect states are introduced within the band gap and greatly increase the carrier concentration. With stronger reduction, intermediate substoichiometric molybdenum oxide phases and eventually metallic $MoO_2$ may form. The electronic properties therefore evolve from wide-gap wide-gap semiconducting $MoO_3$ toward metallic $MoO_2$. The fact that our $MoO_x$ film is highly insulating proves that $x$ is close to 3.

## S3. Nano-Infrared Measurements

Nano-infrared imaging experiments were performed using a scattering-type scanning near-field optical microscope (s-SNOM; Neaspec GmbH) integrated with an atomic force microscope (AFM), enabling simultaneous acquisition of surface topography and near-field optical signals. The AFM was operated in tapping mode with a typical tapping frequency of about 270 kHz and a tapping amplitude of 50 nm. Platinum–iridium-coated silicon probes (NanoWorld AG) with a nominal tip radius of approximately 25 nm were used throughout the measurements. The scattered optical signal was detected using a mercury cadmium telluride (MCT) photodetector (Kolmar Technologies). A pseudo-heterodyne interferometric detection scheme was employed to retrieve both the amplitude ($s$) and phase ($\psi$) of the complex near-field signal. To suppress far-field background contributions, the detected signal was demodulated at the third harmonic of the AFM tapping frequency. Optical excitation was provided by a continuous-wave $CO_2$ laser (Access Laser Co.) with discrete emission wavelengths ranging from 9.2 to 10.8 μm, corresponding to photon energies of 115–135 meV. All nano-infrared measurements were performed under ambient laboratory conditions.

**S4. Quantitative modeling of plasmon interference fringes**

Quantitative analysis of the plasmon interference fringes was performed using a previously established spheroid model (see Ref. 3 in the main text). In this model, the s-SNOM probe is approximated as a conducting spheroid with a total length of $2L$ and a tip-apex radius of curvature $a$. The parameter ($a$) was fixed at 25 nm, consistent with the nominal tip radius specified by the manufacturer, while $L$ was chosen to be much larger than $a$, representing the elongated geometry of the metallic probe. The scattered near-field signal was assumed to be proportional to the induced dipole moment of the tip ($p_z$), along its axis. To simulate the experimentally measured third-harmonic near-field response, $p_z$ was calculated as a function of the instantaneous tip–sample separation throughout the tapping cycle, and the resulting signal was demodulated at the third harmonic of the tapping frequency. Plasmon interference fringe profiles were then obtained by calculating the third-harmonic near-field amplitude at different in-plane tip positions relative to the graphene edge and other scattering boundaries. The graphene response is parameterized by the complex plasmon wavevector $q_p = q_1 + iq_2$, or equivalently by the plasmon wavelength $\lambda_p = 2\pi/q_1$ and damping rate $\gamma_p = q_2/q_1$. By fitting the experimental fringe profiles, as illustrated in Fig. 1e of the main text and Fig. S1, we determine $\lambda_p$ and $\gamma_p$.

**S5. Additional Plasmon Fringe Fitting**

Figure S1 shows experimental fringe profiles (black solid curves) and calculated profiles (red dashed curves) acquired at different excitation energies, $MoO_x$ thicknesses, and gate voltages. The profiles were extracted perpendicular to the sample edges from the datasets summarized in Figs. 2–4 of the main text and are vertically offset for clarity. The calculated profiles were obtained using the quantitative spheroid model described in Section S4. By fitting the experimental profiles, we extracted the plasmon wavelength $\lambda_p$ and damping rate $\gamma_p$. The extracted $\lambda_p$ values were used to construct the experimental plasmon dispersion shown in Fig. 2g of the main text and, with the dispersion calculations described in Section S6, to determine the graphene Fermi energy $E_F$ and carrier density $n$.

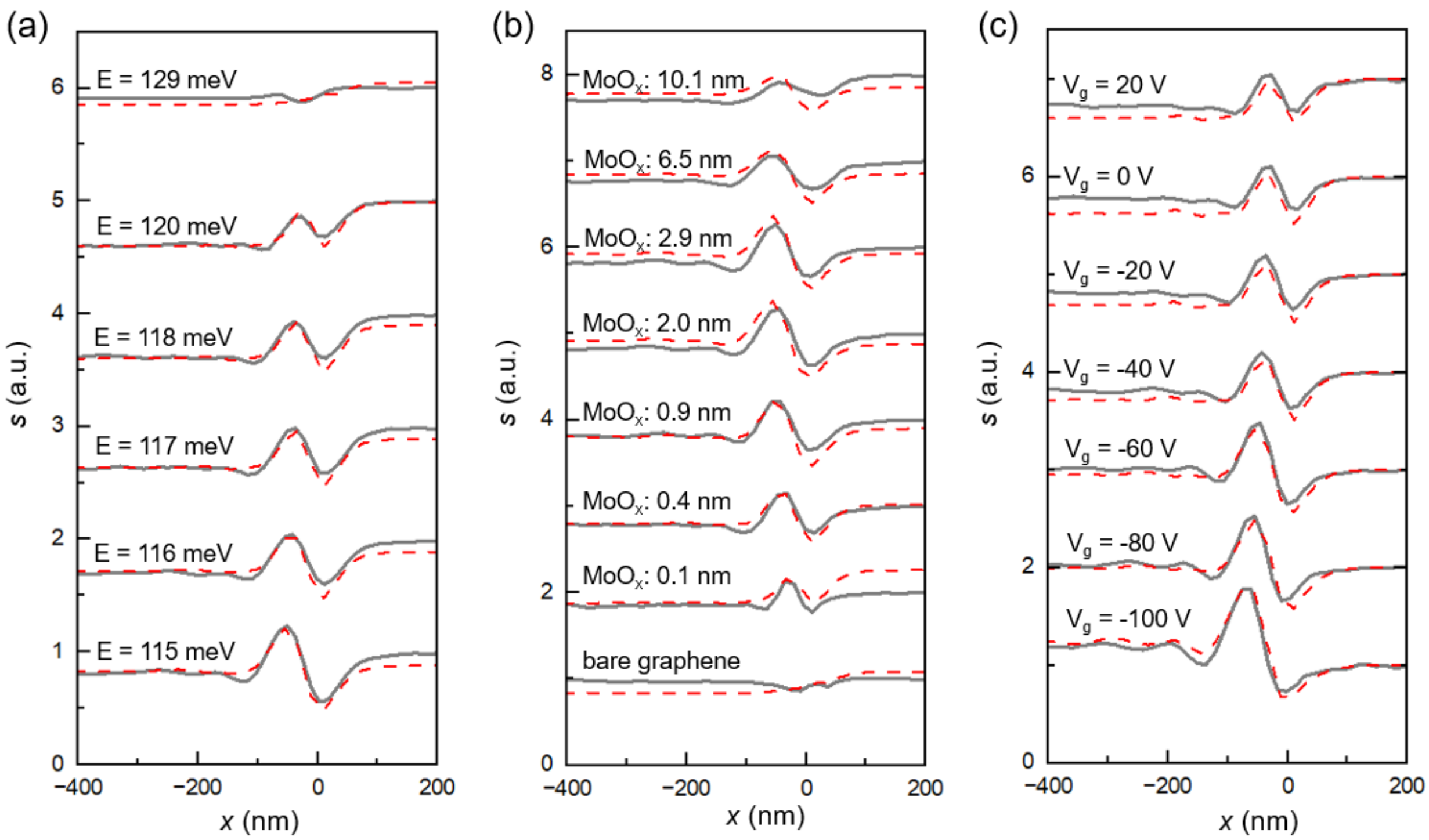

**Fig. S1**. Experimental profiles (black solid curves) and calculated profiles (red dashed curves), vertically offset for clarity. (a) Fringe profiles of the $MoO_x$/graphene heterostructure acquired at different excitation energies. (b) Fringe profiles acquired at different $MoO_x$ thicknesses, including intermediate thicknesses not displayed as images in the main text. (c) Fringe profiles acquired at different back-gate voltages, including intermediate voltages not displayed as images in the main text. The datasets correspond to the measurements summarized in Figs. 2–4 of the main text.

**S6. Calculation of the plasmon dispersion**

The frequency- and momentum-dependent plasmon dispersion colormap shown in Fig. 2g of the main text was obtained by computing the imaginary part of the p-polarized reflection coefficient, $\mathrm{Im}(r_p)$, of the full $MoO_x$/graphene/$SiO_2$/Si multilayer stack. The calculations were performed using the transfer-matrix method, with graphene treated as a two-dimensional conducting sheet. The optical conductivity of graphene was calculated within the random phase approximation [5]. Polaritonic modes appear as peaks in $\mathrm{Im}(r_p)$associated with poles of $r_p$[5–7]. The frequency-dependent permittivity of $MoO_x$ ($x \approx 3$) was adopted from previous infrared studies of isotropic $MoO_3$ films [8,9]. Because the $MoO_x$ layers are ultrathin, the calculated results are not very sensitive to modest variations in their optical constants. The permittivity of Si was set to 11.4 throughout the measured spectral range [10], and the optical constants of $SiO_2$ were adopted from Ref. 11. For each $MoO_x$ deposition thickness $d$, the corresponding oxide thickness was explicitly included in the multilayer model. The graphene Fermi energy $E_F$was varied until the calculated plasmon momentum at the experimental excitation energy matched the measured value $q_p = 2\pi/\lambda_p$. The corresponding graphene carrier density was obtained from$n = E_F^2/(\pi\hbar^2 {v_F}^2)$. This procedure was applied at each $MoO_x$ thickness to determine $E_F(d)$and $n(d)$, producing the thickness-dependent carrier-density data shown in Fig. 3h of the main text. Figure S2 presents representative calculated dispersion colormaps for $E_F = 0.2$, 0.3, and 0.4 eV. At a fixed excitation energy of 115 meV, increasing $E_F$shifts the plasmon mode toward lower momentum, corresponding to a longer plasmon wavelength.

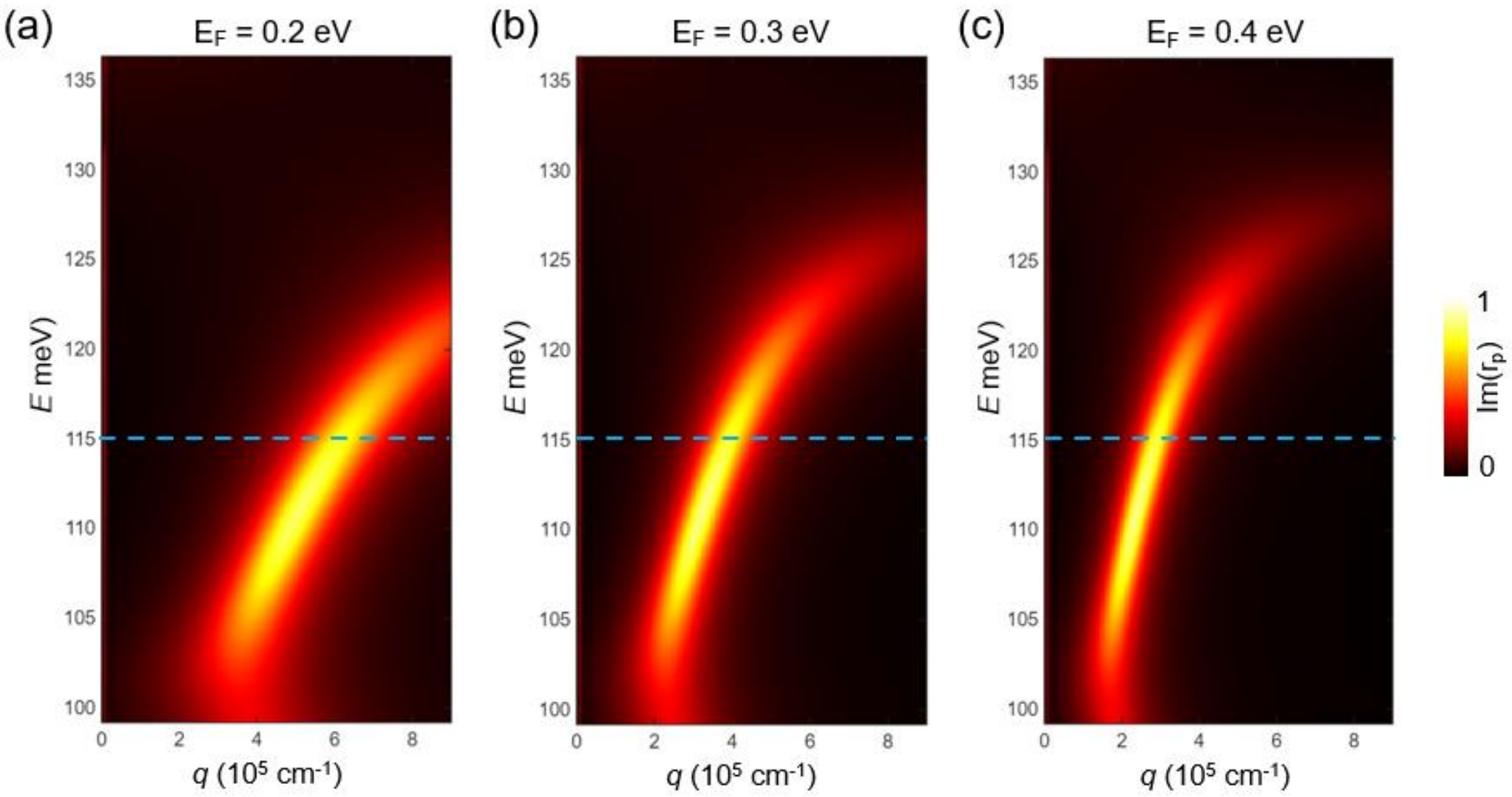

**Fig. S2**. Calculated dispersion colormaps of the $MoO_x$/graphene/$SiO_2$/Si heterostructure with different Fermi energy ($E_F$) of graphene. The blue dashed line marks the IR energy of 115 meV.

**S7. Thickness-dependent phenomenological model**

The phenomenological model about thickness dependence used in the main text can be motivated by assuming that the incremental charge-transfer contribution from a differential $MoO_x$ layer decreases exponentially with its distance $z$ from graphene. In the main text, we mainly discuss two types of charge-transfer mechanisms: (1) short-range direct charge transfer due to wavefunction overlapping, and (2) a possible long-range plasmon-assisted charge transfer. For each charge-transfer mechanism (labeled with $i$), we write the local contribution per unit oxide thickness as

$$\rho_i(z) = \frac{A_i}{L_i}\exp\left(-\frac{z}{L_i}\right),$$

where $L_i$ is the effective decay length and $A_i$ is the saturation carrier-density contribution of that channel. For a $MoO_x$ overlayer with total thickness $d$, the accumulated contribution to the graphene carrier density is obtained by integrating over the oxide thickness:

$$\Delta n_i(d) = \int_0^d \rho_i(z)\, dz = A_i\left[1 - \exp\left(-\frac{d}{L_i}\right)\right].$$

If two charge-redistribution processes characterized by different effective decay lengths contribute independently, their accumulated contributions can be added, giving

$$n(d) = n_0 + A_1\left[1 - \exp\left(-\frac{d}{L_1}\right)\right] + A_2\left[1 - \exp\left(-\frac{d}{L_2}\right)\right],$$

where $n_0$ is the carrier density of bare graphene. This expression is the phenomenological two-length-scale model used to fit the thickness-dependent carrier-density data in Fig. 3h of the main text. The two terms represent the integrated contributions of processes with different effective spatial ranges. This derivation provides physical motivation for the functional form but does not assign either term to a unique microscopic charge-transfer mechanism.

**S7. The vertical decay length of graphene plasmons**

As discussed in the main text, one possible origin for the long-distance charge transfer is the plasmon-assisted charge transfer. The efficiency of such a process is expected to be proportional to the total number of plasmons or the plasmon intensity. Its vertical dependence can be estimated from the evanescent decay of the plasmon field. Let the complex in-plane plasmon wavevector be $q_p = q_1 + iq_2$, where $q_1 = 2\pi/\lambda_p$, and let $k_0 = 2\pi/\lambda_0$be the free-space wavevector. The out-of-plane wavevector in $MoO_x$ is

$$k_z = \left(\varepsilon_{\mathrm{MoO}_x} k_0^2 - q_p^2\right)^{1/2}.$$

Because graphene plasmons are strongly confined,

$$q_p \gg \sqrt{\varepsilon_{\mathrm{MoO}_x}}\, k_0,$$

and therefore $k_z \approx i q_p$. The plasmon-field amplitude consequently decays approximately as

$$| E(z) | \propto \exp(-q_p z),$$

and the corresponding plasmon intensity decays as

$$I(z) \propto | E(z) |^2 \propto \exp(-2 q_p z).$$

Defining the intensity decay length $L_I$ by $I(L_I) = I(0)/e$ gives

$$L_I = \frac{1}{2q_p} = \frac{\lambda_p}{4\pi}.$$

For the experimentally measured plasmon wavelengths $\lambda_p \approx$ 190–200 nm, this relation gives $L_I \approx$ 15–16 nm. This value agrees with the fitted decay length for the long-distance charge transfer process in the main text (see Fig. 3h).

**References for the Supporting Information**